# One-step Synthesis of Cubic Gauche Polymeric Nitrogen with High Yield Just by Heating

Liangfei Wu(吴良飞)[1‡], Yuxuan Xu(徐宇轩)[1,2‡], Guo Chen(陈果)[1,2], Junfeng Ding(丁俊峰)[1,2], Ming Li(李明)[1,2*], Zhi Zeng(曾雉)[1,2], and Xianlong Wang(王贤龙)[1,2*]

[1]Key Laboratory of Materials Physics, Institute of Solid State Physics, HFIPS, Chinese Academy of Sciences, Hefei 230031, China;
[2]University of Science and Technology of China, Hefei 230026, China;
[*]Corresponding author. Email: liming@issp.ac.cn; xlwang@theory.issp.ac.cn
[‡]These authors contributed equally to this work.

**Abstract**

A high-efficient one-step synthesis of cubic gauche polymeric nitrogen was developed just by thermal treatment of $KN_3$ powders. The Raman and infrared spectra confirm the formation of polymeric nitrogen networks. Thermogravimetric differential scanning calorimeter measurements show that the content of cubic gauche polymeric nitrogen is as high as 1.5 wt% with high thermal stability, which is the highest content value so far.

## 1. Introduction

Polymeric nitrogen is characterized by its N-N single or N=N double bonds, when they transform into nitrogen gas containing N≡N triple bonds, the process releases a vast amount of energy. The most attractive polymeric nitrogen is the cubic gauche nitrogen (cg-N), which consists only N-N single bonds with similar crystal structure of diamond [1–4]. After cg-N was successfully synthesized at 110 GPa, LP-N, HLP-N, and BP-N were discovered successively under pressures of 126 GPa, 244 GPa, and 146 GPa, respectively [5–10]. However, the extreme high-temperature and high-pressure conditions required for these syntheses have significantly hindered their practical applications.

In recent years, researchers have begun exploring more moderate synthesis methods, particularly the preparation of cg-N under ambient conditions using plasma-enhanced chemical vapor deposition (PECVD) technology [11–16]. Since 2020, we started to work on this field [17]. Most recently, we reported that KN₃, as a precursor, is more favorable than NaN₃ for the synthesis of cg-N, and free-standing cg-N was successfully synthesized at ambient conditions, which shows high thermal stability up to 760 K and related explosion parameters were reported for the first time [16]. In this work, we report a one-step synthesis of cg-N just by thermal treatment, which is the simplest way to date, and the highest cg-N content is realized.

## 2. Results and discussion

The detail synthesis process and measurement results are shown in the following. The ordered KN₃ powder was heated at temperatures between 150 and 300 °C for 0.5~3 h in vacuum or protection gas. The raw KN₃ is white powder (Fig. 1a) and the color is changed into green after thermal treatment (Fig. 1b). We also compressed KN₃ powder into a white pellet (Fig. 1c) and then the same thermal treatment process was applied. As shown in Fig. 1d, it could be seen that after thermal treatment the surface and interior of the pellet turned green and the center of the interior showed a dark green color, indicating that the thermal treatment can efficiently induce bulk reaction with high yield compared to the surface penetration of PECVD.

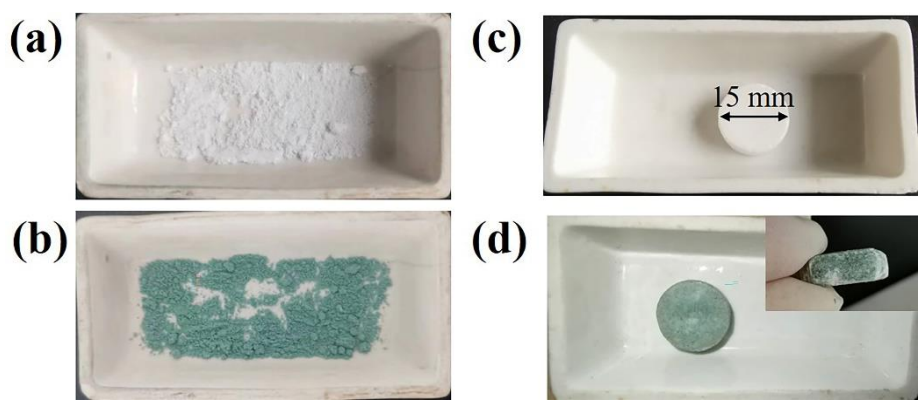

**Fig. 1.** Color changes of different forms of KN₃ before and after heating in a tube furnace. (a) Powder form of KN₃ before reaction. (b) Powder form of KN₃ after reaction. (c) Pellet form of KN₃ before reaction. (d) Pellet form of KN₃ after reaction. The inset image in Fig. 1d shows the cross-sectional view of the pellet sample.

Fig. 2a shows the FTIR spectral profile of the sample obtained by thermal treatment. The characteristic peaks of KN₃ are observed at 640 cm⁻¹ for the bending $v_2$ mode and at 2100 cm⁻¹ for the antisymmetric stretching $v_3$ mode of azide ions. The sample shows a new low-intensity vibrational peak at 883 cm$^{-1}$ after thermal treatment, which corresponds to the T(TO) symmetry vibration of cg-N at ambient temperature and pressure. This is in agreement with the results of infrared spectra measured by Benchafia et al [11]. using PECVD treatment of NaN$_3$, and is consistent with the experimental results of our previous work [16].

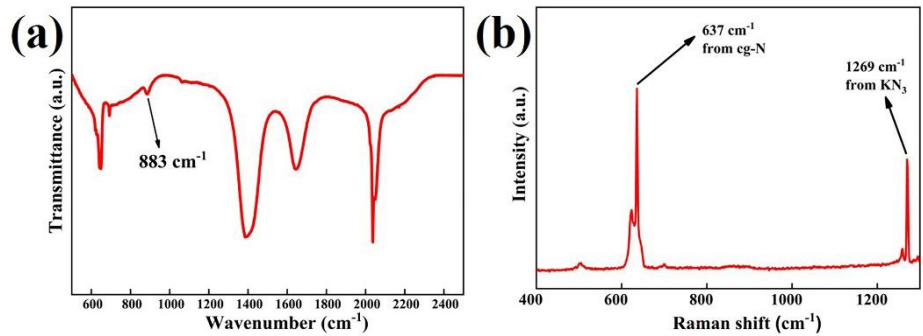

**Fig. 2.** (a) FTIR and (b) Raman spectra of KN$_3$ after the thermal treatment.

Fig. 2b shows the Raman spectral profile of cg-N obtained using thermal treatment. The characteristic peak at 1269 cm$^{-1}$ corresponds to the 2$\nu_2$ mode of the azide ion. A new characteristic peak at 637 cm$^{-1}$ was observed after thermal treatment. This is in agreement with the results measured by Benchafia et al [11]. and with Caracas' density-functional theory calculations of the peak positions and intensities of Raman-activated modes at zero pressure for the cg-N phase [18], as well as the experimental results of our previous work [16]. The results above suggest the successful synthesis of cg-N from KN$_3$ using one-step thermal treatment.

Thermogravimetric differential scanning calorimeter (TG-DSC) analysis of the KN$_3$ after thermal treatment was carried out in a flowing N$_2$ atmosphere (Fig. 3). The DSC curve shows KN$_3$ undergoes a phase transition from the solid state to the liquid state in the temperature range from 325 to 350 °C and the complete decomposition temperature of KN$_3$ is above 500 °C, while the endothermic DSC peak at 448 °C is owing to the decomposition of cg-N. These results were consistent with the experimental results of our previous work by using PECVD treatment.

Importantly, a clear weight loss of ~1.5% at 448 °C in the TG curve corresponds to the decomposition of cg-N, indicating that the cg-N content in this one-step synthesis is ~1.5 wt%, which is two times more than that of our previous work by using PECVD treatment [16]. This is the highest content reported so far, suggesting that our one-step method is high-efficient to obtain cg-N.

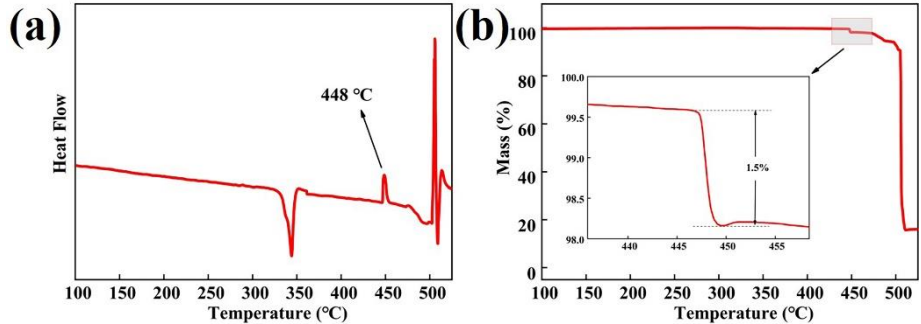

**Fig. 3.** The (a) DSC and (b) TG curves of $KN_3$ after the thermal treatment.

Since we observed the transformation from $KN_3$ into cg-N by PECVD [16], we started to understand the reaction mechanism. First of all, we should answer the question whether potassium exists after the formation of cg-N. Therefore, a small amount of deionized water was added into the tube containing $KN_3$. As shown in Fig. 4b, the raw $KN_3$ powder (Fig. 4a) dissolve immediately and no bubble appears. However, a large number of bubbles were generated in the case of $KN_3$ containing cg-N (Fig. 4d). The gas chromatographic analysis demonstrates the presence of $H_2$, $N_2$ and $O_2$ (Fig. 4e). In this case, $H_2$ can only be generated from the reaction between metal potassium and water. Thus, it can be speculated that metal potassium is precipitated from the $KN_3$ during the reaction process. It is noted that $N_2$ and $O_2$ may derive from the air, therefore determining whether the reaction product of cg-N with water contains $N_2$ is difficult at present and further investigation is needed.

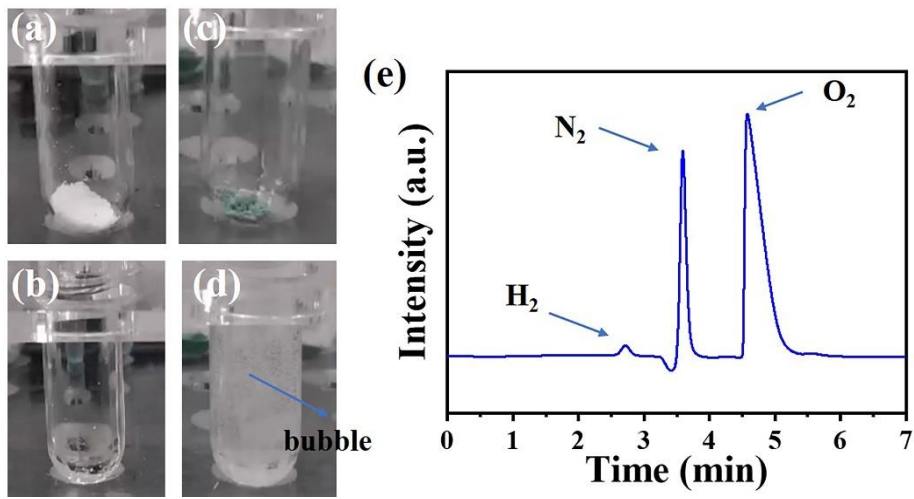

**Fig. 4.** The digital images of KN$_3$ powders (a) before and (b) after adding deionized water. The digital images of thermal treated KN$_3$ (c) before and (d) after adding deionized water. (e) The gas chromatographic analysis of the bubbles.

## 3. Conclusions

In summary, we developed the simplest way to synthesize cg-N with high yield via a one-step scheme. The highest content of cg-N was obtained so far under ambient conditions. The same thermal treatment process was also applied on NaN$_3$, however, the conversion efficiency was lower than that of KN$_3$. We are trying to applied this one-step routine to other azides. Our results will promote the applications of polymeric nitrogen.

**Note**: Just before submitting this manuscript, we notice the work on thermal-chemical route with *multi-steps* to synthesize cg-N [19]. In that work, a solution of azides is prepared first, and then pretreatment under vacuum conditions is used, finally, the obtained precursor is heated at high temperature. Here, we used *one-step* method through heating the primary material under vacuum condition or protection gas environments to get the cg-N.

## Acknowledgements

This work was supported by the CASHIPS Director's Fund (grant no. YZJJ202207-CX, YZJJ202308-TS, and YZJJ-GGZX-2022-01).